# Local noncentrosymmetric structure of $Bi_2Sr_2CaCu_2O_{8+y}$ by X-ray magnetic circular dichroism at Cu K-edge XANES

Andrey A. Ivanov[1], Valentin G. Ivanov[1], Alexey P. Menushenkov[1], Fabrice Wilhelm[2], Andrei Rogalev[2], Alessandro Puri[3], Boby Joseph[4], Wei Xu [5], Augusto Marcelli[6,7], Antonio Bianconi[1,7,8]

[1]*National Research Nuclear University MEPhI (Moscow Engineering Physics Institute), Kashirskoe sh. 31, 115409 Moscow, Russia*
[2] *European Synchrotron Radiation Facility (ESRF), CS40220, F-38043 Grenoble Cedex 9, France*
[3]*CNR-IOM-OGG c/o ESRF LISA CRG, 71 Avenue des Martyrs 38000 Grenoble, France*
[4] *Sincrotrone Elettra, Strada Statale 14 - Km 163,5 Area Science Park, 34149 Basovizza, Trieste, Italy*
[5] *Beijing Synchrotron Radiation Facility Institute of High Energy Physics Chinese Academy of Sciences Beijing, 100049 P. R. China*
[6] *Istituto Nazionale di Fisica Nucleare, Laboratori Nazionali di Frascati, 00044 Frascati (RM), Italy*
[7] *Rome International Centre for Material Science Superstripes, RICMASS, via dei Sabelli 119A, 00185 Rome, Italy*
[8]*Institute of Crystallography, Consiglio Nazionale delle Ricerche, CNR-IC, via Salaria, Km 29.300, 00015 Monterotondo, Roma, Italy*

## Abstract

The two-dimensional $Bi_2Sr_2CaCu_2O_{8+y}$ (Bi2212), the most studied prototype cuprate superconductor, is a lamellar system made of a stack of two-dimensional corrugated $CuO_2$ bilayers separated by $Bi_2O_{2+y}Sr_2O_2$ layers. While the large majority of theories, proposed to interpret unconventional high $T_c$ superconductivity in $Bi_2Sr_2CaCu_2O_{8+y}$, assume a centrosymmetric tetragonal $CuO_2$ lattice for the $[CuO_2]Ca[CuO_2]$ bilayer here we report new compelling results providing evidence for local non-centrosymmetric symmetry at the Cu atom. We have measured polarized Cu *K*-edge XANES (x-ray absorption near edge structure) and the *K*-edge X-ray magnetic circular dichroism (XMCD) of a Bi2212 single crystal near optimum doping. The Cu K edge XMCD signal was measured at ID12 beamline of ESRF with the k-vector of x-ray beam parallel to c-axis i.e. with the electric field of x-ray beam **E**//**ab**, using a 17 T magnetic field parallel to the c-axis of a Bi2212 single crystal. Numerical simulations of the XMCD signal of Bi2212 by multiple scattering theory have shown agreement with the experimental XMCD signal only for the local structure with non-centrosymmetric Bb2b space group of $Bi_2Sr_2CaCu_2O_{8+y}$.

## Introduction

$Bi_2Sr_2CaCu_2O_{8+y}$ (Bi-2212) is considered the prototype of copper-based high-temperature superconductors in which all key features of these systems are clearly manifested. The crystal structure of Bi2212 is made of superconducting $CuO_2$ bilayers which are separated by oxide insulating layers $SrO_2BiO_{2+y}$ layers forming a superlattice of quantum wells with a 3 nm periodicity. While Bi2212 has been object of many experimental investigations spanning about three decades its structure has remained enigmatic. The earliest space-group assignment for the Bi2212 structure was the tetragonal I4/mmm with a=b=0.3817 nm, c= 3.06 nm [1-2] or the orthorhombic Fmmm (a=b=0.54 nm, c=3.08 nm) [3] as expected by the standard BCS theory developed for a homogeneous single band superconductor where the disorder induces a decrease of the critical temperature toward zero. These early reports have been falsified by many experiments which have shown on the contrary that the macroscopic quantum coherence giving 90K superconductivity in Bi2212 emerges in a very complex disordered inhomogeneous phase from atomic scale to micron scale which it an hot topic today after 30 years of research in this field. The refinements of the crystal structure by electron, x-ray and neutron diffraction [4-12] and resonant x-ray scattering at the Cu *K*-edge [13] revealed a monoclinic structure and an incommensurate modulation along the long b axis with period $\lambda/b \sim 4.7$ as shown in Fig. 1.

Two different space groups for the average structures of Bi2212 have been proposed in the literature: first, noncentrosymmetric Bb2b or A2aa space group [5,6] and second, centrosymmetric Bbmb space group [7-13] proposed by teams neglecting the weak reflections probing the non-centric orthorhombic cell, which permits the oxygen in the BiO plane to move off the center of the Bi square approaching within 2.2 Å a pair of Bi atoms [5.6]. The Cu edge resonant elastic x-ray scattering [13] has shown that the Cu plane follows large amplitude incommensurate displacements with different modulations of the Cu and planar oxygen atoms.

After thirty years of intense studies till date there is no consensus on the crystallographic space group of Bi2212. Therefore we have carried out x-ray magnetic circular dichroism (XMCD) measurements at the Cu *K*-edge x-ray absorption near edge structure (XANES) spectra which can select between the two proposed space groups: non-centrosymmetric versus the centrosymmetric.

Cu *K*-edge XANES spectra in a range of 30 eV above the absorption threshold are determined by the multiple scattering resonances of the excited photoelectron with *p* orbital symmetry excited from the 1*s* core level of the absorbing atom [14-19]. It was shown in the seventies that the photoelectron with kinetic energy in the range 10-50 eV has a very short lifetime, short mean free path and strong scattering probability by neighbor atoms therefore at the multiple scattering resonant energy it is confined in a nanoscale cluster [20].

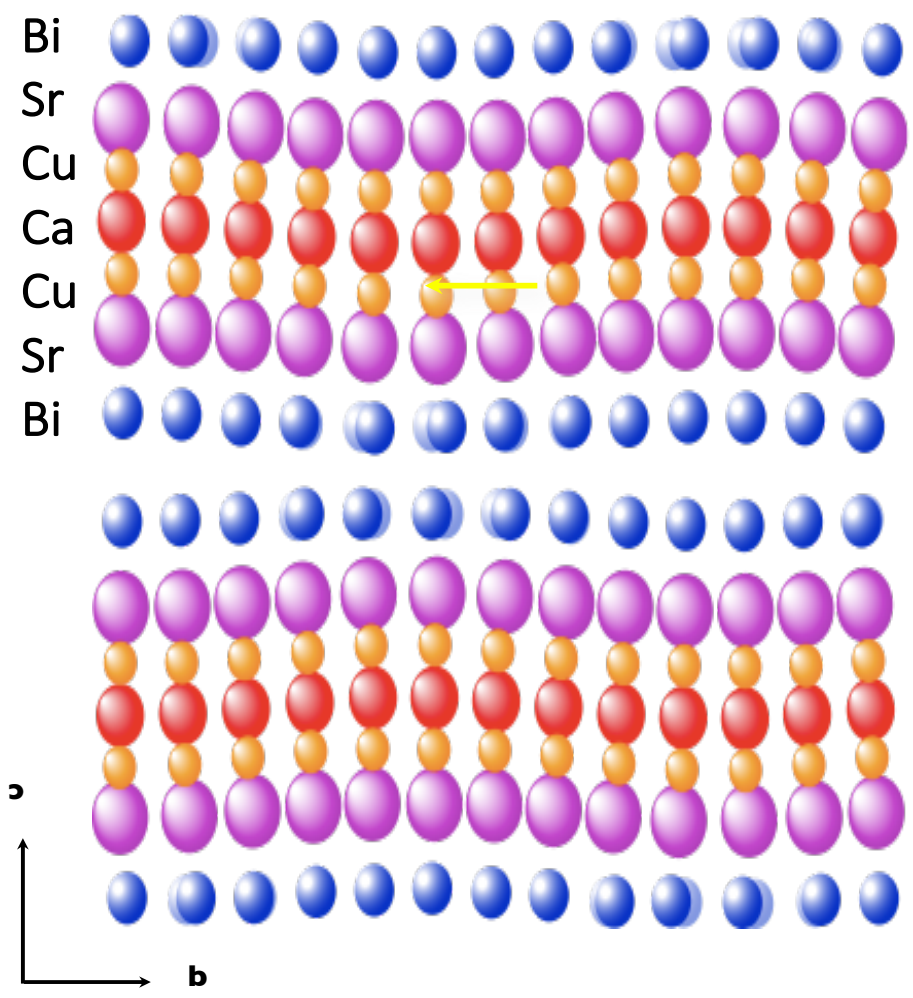

**Figure 1.** The structure of the unit cell of the Bi 2212 crystal plotted in the **bc** plane. The two blocks of thickness c/2 are kept together by van der Waals interactions between the bismuth layers forming the unit cell. The layers have the 1D structural corrugation driven by the misfit strain between the Bi-Sr spacer layers and the metallic Cu-Ca-Cu bilayers. Oxygen is not shown for simplicity.

X-ray magnetic circular dichroism (XMCD) is a very powerful element-selective technique which makes possible to study magnetic properties of materials at microscopic scale [21-27]. The XMCD spectrum is a difference between x-ray absorption near edge structure (XANES) spectra recorded at the selected x-ray absorption edge of a specific atomic element collected for two opposite helicities of the incident beam [21-27]. This effect may be due to intrinsic properties of a system, such as noncentrosymmetric average structure, sawtooth shape of the charge density, or orbital density wave in the presence of a strong external magnetic field, which orients the magnetic moments inside the sample.

The technique is based on the fact that the absorption coefficient of the material under study depends on the helicity of the absorbed light. As a rule we have registered two sets of XMCD spectra for an external magnetic field being parallel or antiparallel to the x-ray beam to rule out possible instrumental artifacts since the variation of the XANES spectrum for the two opposite helicities of the incident beam is very small. A previous investigation of the XNCD of a cuprate superconductor has been reported in Ref. [28] but the results have been questioned by theoretical calculations of Norman [29]. Moreover Di Matteo and Varma [30,31] provided calculated XNCD and XMCD spectra for possible signatures of time-reversal breaking, the chiral order and the pseudogap phase.

XMCD at the Cu $L_{2,3}$-edges of x-ray absorption spectra in both undoped $La_2CuO_4$ and doped $YBa_2Cu_3O_7$ films with $T_c$=90K have been reported [32]. The photon beam was normal to the sample surface and parallel to the c axis and to the external magnetic field. At the $L_2$-edge and $L_3$-edge x-ray absorption process a 2p core electron is excited to 3d empty states of Cu with different spin polarization [33-35] therefore it is the ideal tool to investigate the spin polarization of 3d orbitals. The origin of the detected dichroism [32] could not be explained by the authors through a purely paramagnetic canting of the spins induced by the field, which would require at least a ten times higher magnetic fields to give the measured dichroic signal in the c direction.

We report here measurements of x-ray magnetic circular dichroism (XMCD) at Cu *K*-edge of the $Bi_2Sr_2CaCu_2O_{8+x}$ (Bi2212), a sample close to optimal doping with $T_c$= 87K. $Bi_2Sr_2CaCu_2O_{8+x}$ (Bi2212) is a very complex material with a misfit strain between the metallic atomic layers and the spacer layers [36-39], the short range charge density wave CDW puddles [40-44] and the nanoscale phase separation [45-53]. These interesting phenomena provide a really complex landscape of unconventional superconductivity in metallic 2D layers coupled by Josephson junctions [54-57] in Bi2212.

The metallic 2D layers of Bi2212 are decorated by 1D quantum wires [45,46] arising from the incommensurate modulation.

The misfit strain modulation and the charge density wave modulation induce local lattice fluctuations revealed by anisotropic response of photoemission [58,59] and by anisotropic x-ray absorption probing the local structure [60,61]. The $CuO_2$ lattice fluctuations could be driven by sawtooth modulations of the oxygens in the Bi2O2 slabs [62-67] which could produce a noncentrosymmetric superconducting phase in the $CuO_2$ planes. To study such lattice distortions one can use XMCD technique which is the ideal tool to detect the local noncentrosymmetric structure in layered materials [68-70].

There is a large interest today in superconductivity in noncentrosymmetric materials [71-73], where standard BCS approximations fail, giving unconventional non-BCS scenarios. Noncentrosymmetric superconductivity has been observed mostly in heavy fermions intermetallic noncentrosymmetric structures however a *local noncentrosymmetric* symmetry has been observed in oxides [74-77]. This work shows the *local noncentrosymmetric* Cu site in the case of Bi2212 complex 2D quantum matter opening the possibility of local *noncentrosymmetric condensates* where local *Rashba spin–orbit* coupling should play a subtle role.

## Materials and Methods

Here we have investigated $Bi_2Sr_2CaCu_2O_{8+x}$ single crystals grown by traveling-solvent-floating-zone method. Lamellar samples of mm thickness and lateral dimensions of few $mm^2$ with the orientation of the surface plane (001) have been selected. The single crystallinity of the samples was confirmed by XRD analysis. The oxygen content is hard to determine directly in $Bi_2Sr_2CaCu_2O_{8+y}$ however magnetic susceptibility measurements have shown that the temperature of superconducting transition of the selected sample, is 87 K which corresponds to doping close to the optimum doping.
The studies were performed at the ESRF beamline ID12 (Grenoble, France). A single-crystal sample was attached on a cold finger of a helium continuous flow cryostat and mounted so that the incident x-ray beam was perpendicular to the sample surface in order to have the wave vector parallel to the crystallographic axis c. The size of the beam at the sample was 0.4x0.3 $mm^2$. A magnetic field of 17 T was applied parallel to the incident beam, and measurements were made for two opposite directions of the magnetic field. The XANES spectra were recorded in the fluorescence detection mode at the *K* edge of Cu. The measurements were carried out at sample temperatures of 50K, 80K, 100K and 200K. The XMCD spectrum is the difference between two XANES spectra collected with two opposite circular polarizations of the incident beam. In the experiment, as a rule, such a spectrum is obtained by consecutive switching the polarization of the incident beam (+) (-), obtaining a XMCD spectrum from each pair of XANES spectra and repeating the procedure until the required signal-to-noise ratio is obtained. Since the magnitude of the expected XMCD signal was not more than $10^{-4}$, another approach was used to reduce various instabilities and drifts, which may not be eliminated by averaging. Each individual XMCD spectrum was obtained by processing a group of four successively taken XANES spectra, for which the polarization of the incident beam varied as (+) (-) (-) (+), a magnetic field was also changed and then this acquisition sequence was repeated several times for two opposite magnetic field directions.

As the result of the accumulation of several such spectra (their number varied from 40 to 100), the measurement error was lower than $2\times10^{-5}$.

## Results and discussion

The polarized E//ab Cu *K*-edge XANES spectra of the Bi2212 crystal are shown in Fig.3. XANES spectra probe the structure of a nanoscale cluster centered at the absorbing Cu atom [20] shown in Fig. 2 via full multiple scattering of the excited photoelectron with *p*-orbital symmetry. The present experimental data are in good agreement with previous experiments [20]. The weak shoulder in the pre-edge range of the Cu *K*-edge XANES spectrum probes the unoccupied Cu($\varepsilon p$) states near the Fermi level [33].

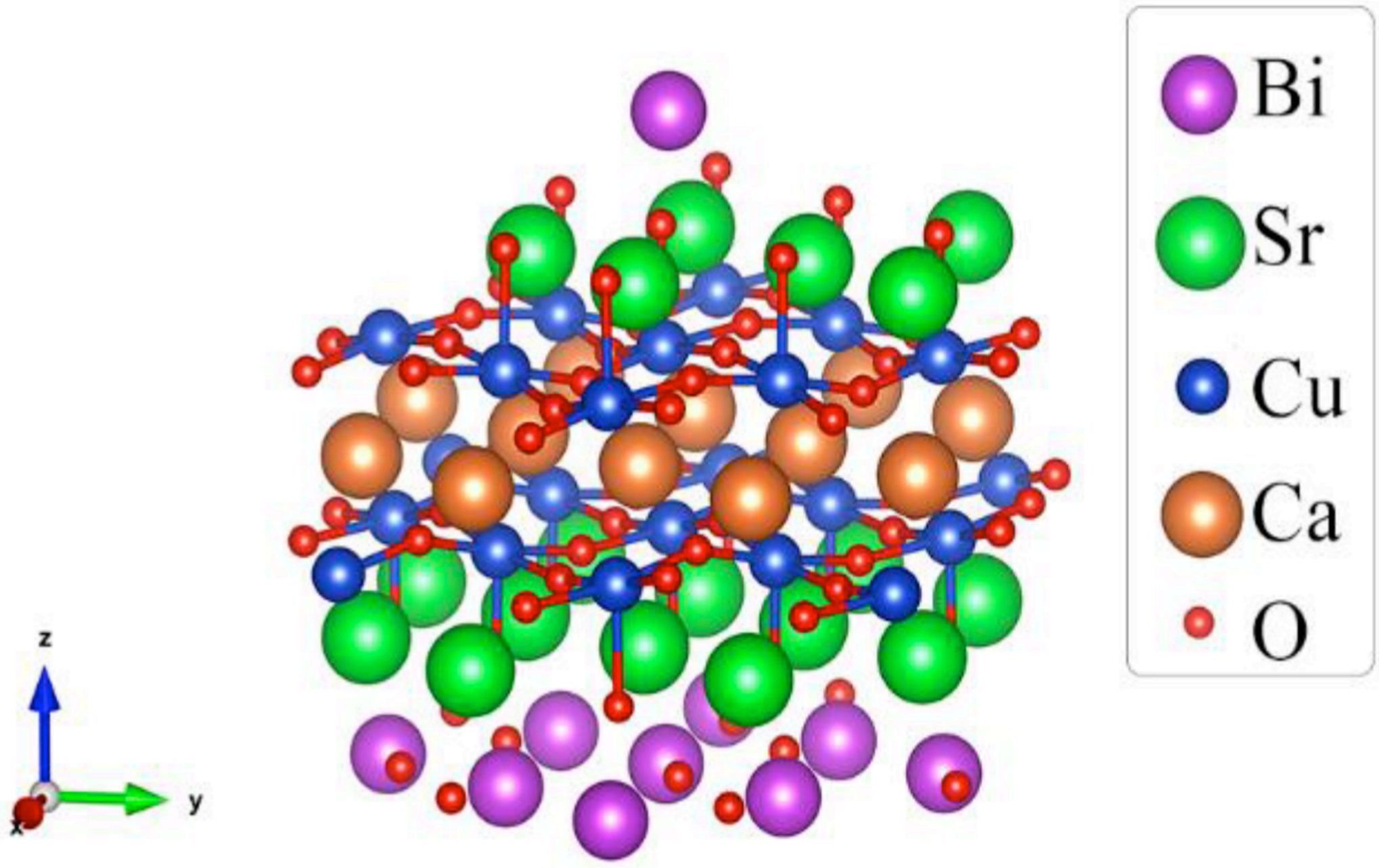

**Figure 2.** The local nanoscale structure of Bi 2212 centered at the Cu site of the atomic cluster probed by Cu *K*-edge XANES. This nanoscale cluster is defined as the spatial range where the excited photoelectrons with kinetic energy 10 eV <E<50 eV is confined within its short lifetime. Only atoms in this cluster, contribute to the multiple scattering resonances (MSR) giving peaks in the XANES spectrum. The cluster is made of the central absorbing Cu atom, its nearest 4 oxygen atoms in the first shell and Cu, Sr, Ca, Bi, O atoms in the further neighboring shells.

The very weak intensity as in other Cu K-edge spectra of cuprate superconductors [18] indicates the lack of itinerant Cu($\varepsilon p$) states at the Fermi level and it is due to the quadrupole transition Cu(1*s*) to Cu(3*d*) which is very weak. Experimental Cu *K*-edge XANES and XMCD spectra presented in Fig. 3 have been taken at 80 K, 100 K and 200 K.

.

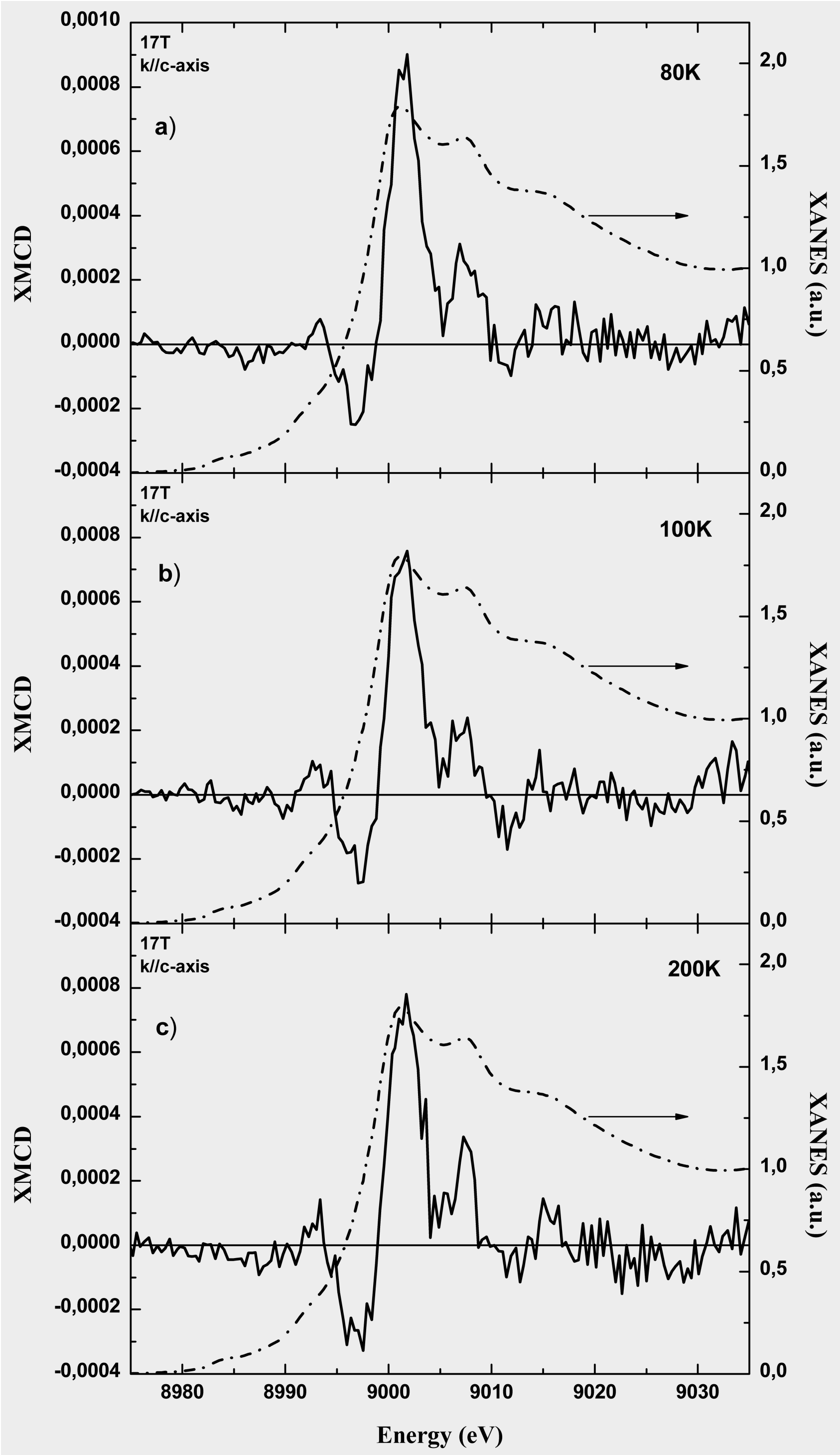

**Figure 3.** Cu K-edge XMCD signal of $Bi_2Sr_2CaCu_2O_{8+y}$ (Bi2212) (solid line) in a magnetic field B of $\mu_0$H=17 T aligned along the c-axis (i.e., **B**//**c**) and polarized **E**//**ab** Cu *K*-edge XANES (point – dashed line) measured at three different temperatures: a) 80K; b) 100K; c) 200K. At 80K the high magnetic field drives the 87K superconducting phase into the normal phase.

The measurements show that though rather small, a clear and reproducible XMCD signal. Its magnitude ~$8\times10^{-4}$ a.u. with respect the unitary edge jump is well above the detectivity limit. It is noteworthy that its energy profile does not coincide with the derivative of the XANES spectrum.

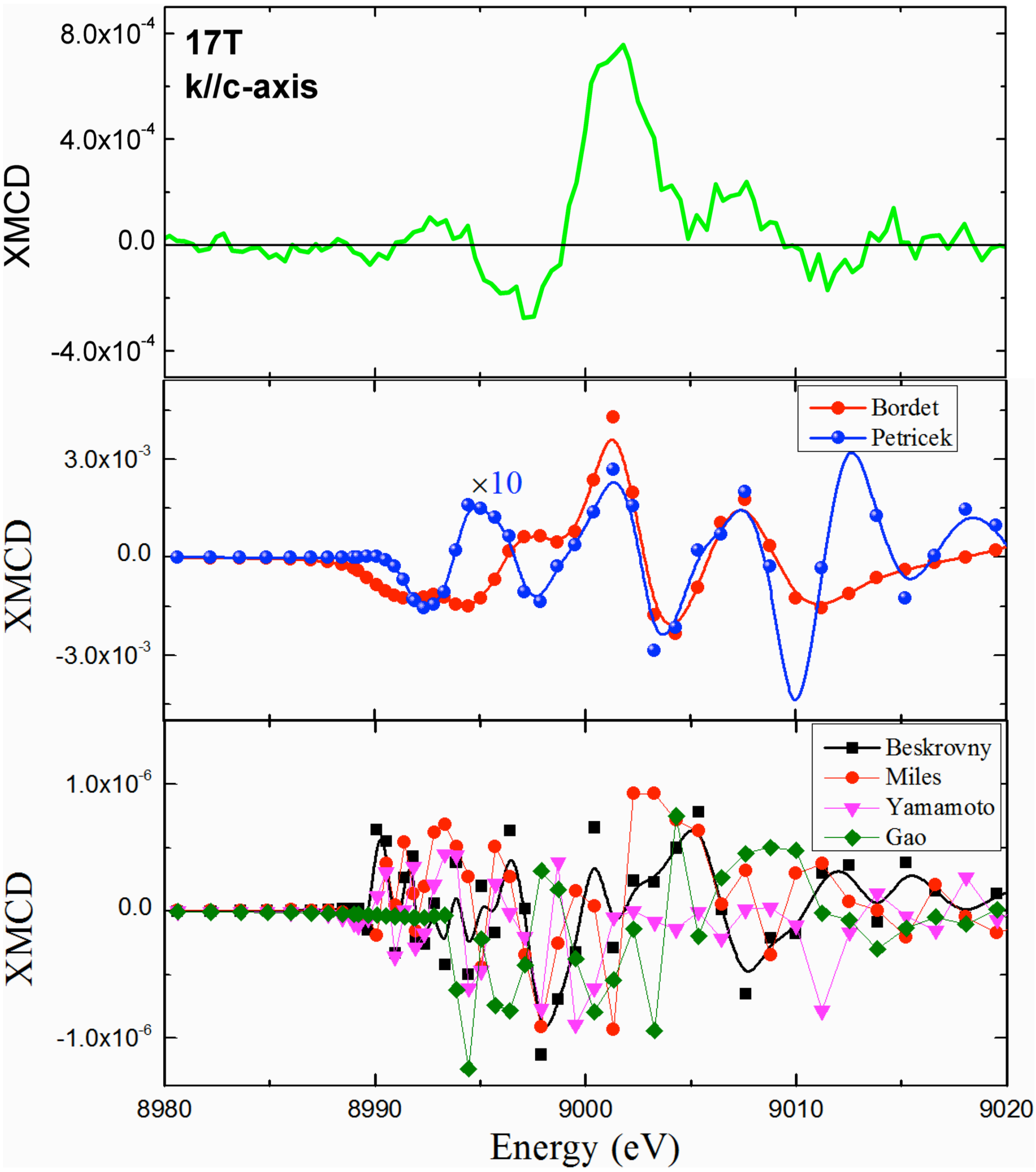

**Figure 4.** *Upper panel*: experimental XMCD spectrum at Cu *K*-edge polarized XANES of Bi2212 taken at 100K in magnetic field of $\mu_0$H=17T; *Center panel*: calculated XMCD spectra in the Cu XANES spectra using the non-centrosymmetric clusters using the crystallographic data measured by Bordet et al. [5] (red points) and by Petricek et al. [6] (blue points); *Bottom panel*: simulated XMCD spectra calculated using the centrosymmetric crystallographic structures reported by Gao et al. [7], Beskrovnyi et al. [9], Yamamoto et al. [10] and Miles et al. [12]. The signal for the centrosymmetric structures is 3 orders of magnitude smaller than that of the non-centrosymmetric structures. Therefore the XMCD signal predicted for a centrosymmetric is expected hidden in the background noise.

The experimental data have been interpreted by theoretical calculations performed using FEFF9 code, which uses full multiple scattering theory (FMST) and self-consistent field method (SCF) with the muffin-tin approximation for the atomic potential. The Hedin-Lundqvist exchange-correlation potential was selected for correction. The radii of atomic clusters around the central absorber atom were 10 and 5 Å for FMST and SCF respectively.

The atomic cluster shown in Fig. 2 was large enough to achieve good convergence. The calculations were made for several different variants of Bi2212 structures, including both *Bb2b* and *Bbmb*. Theoretical XANES spectra have been calculated starting from structural data for of the noncentrosymmetric crystallographic structures of Bordet et al. [5] and Petricek et al. [6], and for the centrosymmetric crystallographic structures reported by Gao et al. [7], Beskrovnyi et al. [9], Yamamoto et al. [10] Kan and Moss [11] and Miles et al. [12]. Theoretical XANES spectra starting from different crystallographic structures [5-13] grab the key experimental features with small differences but there is no evident correlation between the spectrum shape and the space group of the starting structure. On the contrary the calculated XMCD reported in Fig. 4 show a dramatic sensitivity to the symmetry of the crystal space group. Indeed, for a nonzero XMCD signal related with a magnetic signal in the E1-E1 channel, parity even, [31] we need a lattice with a magnetic structure with a net magnetization. Therefore we expect no XMCD signal for structural clusters with centrosymmetric space group due to symmetry considerations and a non zero signal for local noncentrosymmetric clusters.

Our calculations confirmed that this is the case for Bi2212. The simulated XMCD spectra for different diffraction data refinements together with the experimental XMCD spectrum measured at 100K in a magnetic field of $\mu_0$H=17 T are presented in Fig 4.

According to simulations shown in Fig. 4 (lowest panel), the centrosymmetric *Bbmb* refinements give the magnitude of XMCD signal of the order of $10^{-6}$ shown in Fig. 4 (lowest panel), almost three orders smaller than simulated XMCD for noncentrosymmetric *Bb2b* refinements (Fig. 4 - central panel). The XMCD signal given by the noncentrosymmetric structure found by Petricek is ten times smaller than that given by Bordet et al. due to smaller noncentrosymmetric atomic displacements therefore it has been increased by a factor 10.

Thus based on experimental data and in comparison with the results of simulation we can state that the Bi2212 structure belongs to noncentrosymmetric *Bb2b* space group.

If we compare then the experimental XMCD spectrum with simulated ones based on different *Bb2b* refinements we point out that in spite of a discrepancy in the magnitude, the version of structural data given by Bordet et al. [6] provides a better agreement with the energy profile of the experimental spectrum. The issue that remains unaddressed is the nature of magnetic moments that are oriented in the applied external magnetic field, and deserves further work taking into account the very complex multiscale phase separation with multiple lattice and electronic components [51-53] driven by the proximity to Lifshitz electronic topological transition [49-51] in a strongly correlated quasi 2D metal.

**Conclusions**

We have reported a study of x-ray magnetic circular dichroism (XMCD) at the *K* edge of Cu in the high-temperature superconductor $Bi_2Sr_2CaCu_2O_{8+x}$ (Bi2212). The measurements revealed a XMCD signal with magnitude about $8\times10^{-4}$ a.u. in a magnetic field of 17 T. Numerical simulations of the XMCD signal for a series of BSCO lattice distortions showed that only the local noncentrosymmetric structure is the responsible of the non-zero XMCD signal in Bi2212 for our measurement layout. Therefore, this experiment supports the early assignment of the non-centrosymmetric *Bb2b* space group for the Bi2212 crystal structure. Moreover, we propose that the local noncentrosymmetric structure in the Cu site could be driven by sawtooth modulations of the oxygens in the $Bi_2O_2$ slabs [62-67] which will result in a noncentrosymmetric superconducting phase in the 2D $CuO_2$ layers Finally, taking into account that Bi2212 is characterized by a local non-centrosymmetric structure like in few other oxides [75-77], we suggest to confirm these results by spin resolved ARPES experiments and by further experiments looking for the unique features of superconductivity in non-centrosymmetric systems [78-87] on Bi2212 samples.

**Acknowledgements**

The experiment has been supported by superstripes-onlus. The experiments were performed on beamline *ID12* at the European Synchrotron Radiation Facility (ESRF), Grenoble, France. We are grateful to ESRF for beam time allocation, traveling support and ESRF staff for providing assistance in using beamline. Wei Xu acknowledges the financial support from NSFC (Grant No. U1532128) and LNF from the framework of INFN&IHEP collaboration.